\documentclass{emulateapj}
\usepackage{apjfonts}
\usepackage{graphicx}

\usepackage{amsmath}

\shorttitle{Spectral Modification of Hadronic Gamma Rays}
\shortauthors{T. INOUE et al.}

\begin{document}

\title{
Bell-Instability-Mediated Spectral Modulation of Hadronic Gamma Rays from a Supernova Remnant Interacting with a Molecular Cloud}
\author{Tsuyoshi Inoue\altaffilmark{1}}
\altaffiltext{1}{Department of Physics, Graduate School of Science, Nagoya University, Furo-cho, Chikusa-ku, Nagoya 464-8602, Japan; tsuyoshi.inoue@nagoya-u.jp}

\begin{abstract}
Supernova remnants (SNRs) are believed to be the site of galactic cosmic-ray acceleration.
However, the details of the cosmic-ray acceleration are still not well understood.
Gamma ray observation is a promising method to study cosmic-ray acceleration in the SNRs, because a hadronic gamma ray can trace high energy cosmic-rays above $\sim$GeV energy.
Conventional theory predicts that the hadronic gamma ray shows a flat $\nu\,F_{\nu}$ spectrum from the pion-creation threshold energy to the maximum energy of diffusive shock acceleration.
In this paper, by employing numerical simulations that solve a hybrid system of the magnetohydrodynamics of a molecular cloud and diffusive propagation of cosmic-rays, 
we demonstrate that the hadronic gamma ray spectrum can be harder than the conventional one and that the modulated spectrum becomes consistent with observations.
The modification mechanism is explained as follows:
The cosmic-rays accelerated at the supernova blast wave shock propagate into a clump of a molecular cloud.
The cosmic-ray streaming in the cloud induces the so-called Bell instability that induces Alfv\'en waves in the cloud.
The induced magnetic field fluctuations prevent further cosmic-ray incursion by diminishing the diffusion coefficient for the cosmic-rays below $\sim1$ TeV energy.
 This mechanism reinforces recent claims of a similar spectral modification by magnetic field amplification around a molecular cloud by Inoue et al.~(2012) and Celli et al.~(2018).
\end{abstract}

\keywords{waves --- methods: numerical --- ISM: supernova remnants --- gamma rays: ISM}

\section{Introduction}
Supernova remnants (SNRs) are believed to be the site of the galactic cosmic-ray acceleration by diffusive shock acceleration (DSA; Bell 1978, Blandford \& Ostriker 1978, Blandford \& Eichler 1987).
However, it is still not clear how much of the supernova energy is consumed by particle acceleration and whether the SNRs can accelerate particles up to the knee energy ($\sim 10^{15.5}$ eV).
Gamma ray observation is a promising method to understand cosmic-ray acceleration in SNRs, because hadronic gamma rays, which are generated by the decay of neutral pions created by the collisions of interstellar and cosmic-ray protons, can trace high-energy cosmic-rays above $\sim$ GeV energy (Aharonian et al.~2008; Abdo et al.~2009; Fukui et al.~2012; Ackermann et al.~2013; Acero et al.~2017).

Conventional theory, which assumes a spatially homogeneous distribution of both cosmic-rays and the interstellar medium (ISM), predicts that the hadronic gamma rays show a flat $\nu\,F_{\nu}$ spectrum from the pion-creation threshold energy ($\sim 0.1$ GeV) to the maximum energy achieved by the DSA (Naito \& Takahara 1994; Drury et al.~1994).
Results of gamma ray observations for young SNRs suggest that $\nu F_{\nu}$ spectra are generally harder than the conventional hadronic spectrum and the observed spectra are well explained by the leptonic scenario in which gamma rays are created by inverse Compton scattering of the cosmic microwave background photons by the cosmic-ray electrons (see, e.g., Abdo et al.~2011).

However, if we consider an inhomogeneous ISM, the hadronic spectrum can be harder, because cosmic-rays with different energies can interact with different amounts of the ISM protons (Gabici et al.~2009; Zirakashvili \& Aharonian 2010; Inoue et al.~2012).
In the case of a young SNR, RX J1713.7$-$3946, the SNR blast wave is suggested to be interacting with clumpy molecular clouds (Fukui et al.~2003, 2012; Sano et al.~2010, 2015).
Inoue et al.~(2012) showed that the hadronic gamma ray spectrum from such a SNR can be as hard as the leptonic gamma ray spectrum, if the diffusion coefficient for the cosmic-ray protons is proportional to their gyroradius, which can be expected in a turbulent medium.
Gabici \& Aharonian (2014) calculated the detailed spectrum from a molecular clump that is embedded in SNR RX J1713.7$-$3946 by assuming parameters such as the diffusion coefficient and magnetic field strength in the molecular cloud, and found that the gamma ray spectrum can be well fitted to the observational data by {\it Fermi} and H.E.S.S. observatories\footnote{
In the case of RX J1713.7$-$3946, absence of thermal x-ray radiation is claimed as an evidence of non-interaction with dense gas (Ellison et al.~2010).
Inoue et al.~(2012) proposed that the temperature of shocked dense cloud can be below $\sim$1keV depending on the cloud density, and that the thermal x-ray emission can be minimized.
Whether the thermal x-ray radiation is suppressed even from low density cloud envelope will be studied in our future work.}.

In this line of research, Celli et al.~(2018) recently studied cosmic-ray propagation into a molecular clump that is swept up by an SNR blast wave.
As pointed out by Inoue et al.~(2009, 2010, 2012) and Sano et al.~(2012), the interaction of a dense clump and a shock wave generates turbulence that induces magnetic field amplification around the clump (see also Giacalone \& Jokipii 2007; Guo et al.~2012; Inoue et al.~2013 for a turbulent preshock case).
Hence the turbulent magnetic field is expected to suppress the cosmic-ray propagation into the clump.
Celli et al.~(2018) showed that the magnetic field amplification by the ``shock-cloud" interaction can successfully modulate the resulting gamma ray spectrum when the density contrast between the dense clump and surrounding gas is larger than $10^3$.

Although the turbulent magnetic field induced by the shock-cloud interaction has the potential to change the cosmic-ray propagation into the clump, the scale of the magnetic field fluctuations induced by the interaction would be a sub-parsec scale, $\sim10^{17}$ cm, which is much larger than the gyroscale for a GeV proton $\simeq 10^{12}\,(B/10\mu\mbox{G})^{-1}$ cm.
This indicates that the turbulent magnetic field produced by the shock-cloud interaction might be unavailable as a scatterer for GeV to TeV particles (Roh et al.~2016).
Even in such a case, we can still expect another mechanism to generate turbulent magnetic fields inside the clump, owing to cosmic-ray streaming.
Because the cosmic-rays are not efficiently accelerated inside the cloud due to, e.g., the shock stall, the cloud is irradiated by the cosmic-rays that are accelerated in the diffuse inter-clump/inter-cloud medium.
Thus, the clumps of the cold cloud experience a cosmic-ray streaming when the shock approaches them.
As with the Bell instability at the shock upstream (Bell 2004), the cosmic-ray streaming into the cloud induces a return current in the background medium that can generate a turbulent magnetic field in the cloud.
As we demonstrate below, the scale of the turbulent magnetic field fluctuations is small enough to work as the scatterer for the GeV-TeV protons.

In this paper, to examine the effect of the cosmic-ray streaming into the cloud around a young SNR, we employ one-dimensional numerical simulations in which the dynamics of the background medium is described as the Bell MHD and the cosmic-ray dynamics is described by the diffusion convection equation.
The paper is organized as follows:
In \S 2, we provide the basic equations and numerical setup.
The results of the simulations and their implications for the hadronic gamma ray emissions are shown in \S 4.
The summary and conclusion of this paper are presented in \S 5.

\section{Basic Equations and Numerical Setups}
\subsection{Basic Equations}\label{BE}
We solve a hybrid system of the Bell MHD and a telegrapher-type diffusion convection equations in one dimension (Bell et al. 2013):
\begin{eqnarray}
&& \frac{\partial\,\rho}{\partial t}+\frac{\partial}{\partial x}(\rho\,v_x)=0,\label{eq1}\\
&& \frac{\partial}{\partial t}(\rho\,v_x)+\frac{\partial}{\partial x}(\rho\,v_x^{2}+p+\frac{B_y^2+B_z^2}{8\,\pi})=0,\label{eq2}\\
&& \frac{\partial}{\partial t}(\rho\,v_y)+\frac{\partial}{\partial x}(\rho\,v_x\,v_y-\frac{B_x\,B_y}{4\pi})=-\frac{1}{c}j^{({\rm ret})}_{x}\,B_z,\label{eq3}\\
&& \frac{\partial}{\partial t}(\rho\,v_z)+\frac{\partial}{\partial x}(\rho\,v_x\,v_z-\frac{B_x\,B_z}{4\pi})=\frac{1}{c}j^{({\rm ret})}_{x}\,B_y,\label{eq4}\\
&& \frac{\partial\,\epsilon}{\partial t}+\frac{\partial}{\partial x}\{v_x(\epsilon+p+\frac{B_y^2+B_z^2}{8\,\pi})-B_x\frac{ (B_y\,v_y+B_z\,v_z) }{4\,\pi}\}=0,\label{eq5}\\
&& \epsilon=\frac{p}{\gamma-1}+\frac{1}{2}\rho\,v^2+\frac{B_y^2+B_z^2}{8\pi},\label{eq6}\\
&& \frac{\partial\,B_y}{\partial t}=\frac{\partial}{\partial x}(B_x\,v_y-B_y\,v_x),\label{eq7}\\
&& \frac{\partial\,B_z}{\partial t}=\frac{\partial}{\partial x}(B_x\,v_z-B_z\,v_x),\label{eq8}\\
&& \frac{\partial B_x}{\partial x}=0,\label{eq9}\\
&& \frac{\partial F_0(x,p)}{\partial t}+\frac{\partial}{\partial x}(v_x\,F_0(x,p))-\frac{1}{3}\frac{\partial\,v_x}{\partial x}\frac{\partial\,F_0(x,p)}{\partial \ln p}=-\frac{c}{3}\frac{\partial\,F_1(x,p)}{\partial x},\label{DC1},\\
&& \frac{\partial F_1(x,p)}{\partial t}+\frac{\partial}{\partial x}(v_x\,F_1(x,p))=-c\frac{\partial\,F_0(x,p)}{\partial x}-\frac{c^2}{3\,\kappa(p,{\bf B})}F_1(x,p),\label{DC2}
\end{eqnarray}
where $j^{({\rm ret})}_x$ is the return current density induced by the cosmic-ray streaming current, i.e., $j^{({\rm ret})}_x=-j^{(\rm cr)}_x$, and $\kappa(p,\vec{B})$ is the diffusion coefficient, which generally depends on the momentum of the cosmic-rays $p$ and the magnetic field.
Eqs.~(\ref{DC1}) and (\ref{DC2}) constitute the diffusion convection equation, where $f_0(x,p)=F_0(x,p)/p^3$ is the isotropic component of the cosmic-ray distribution function and $f_1(x,p)=F_1(x,p)/p^3$ is the anisotropic component so that the distribution function is given by $f(x,\vec{p})=f_0(x,p)+(p_x/p)\,f_1(x,p)$ (see Bell et al.~2013 for higher order equations\footnote{
Eqs.~(\ref{DC1}) and (\ref{DC2}) are from eqs.~(11a) and (11b) of Bell et al.~(2013), where we neglect quadrupole term $f_{i\,j}$ and also $f_y$ and $f_z$ terms.
Due to the omission of $f_y$ and $f_z$, the $y$ and $z$ components of the cosmic-ray current ($j_y, j_z$) are always set to be null.
$j_y$ and $j_z$ are induced when cosmic-rays, whose gyro-radius is smaller than the wave-length of the magnetic field disturbance, stream along the disturbed field.
The current carried by these cosmic-rays with small gyro-radius does not contribute to the growth of the Bell instability.
This is the reason why we limit the integration range of particle momentum in the current calculation given by eq.~(\ref{jcr2}).
}).
The reason why we use the telegrapher-type diffusion convection equations will be explained in \S {\ref{adv}}.
One can easily confirm that these two equations recover the usual diffusion convection equation derived by Skilling (1975), if we take the limit $c\rightarrow\infty$.
Further, we see that when the cosmic-rays stream freely ($\kappa\rightarrow\infty$), the streaming speed becomes $c/\sqrt{3}$, i.e., the free propagation velocity for isotropic cosmic-rays.

The total cosmic-ray current density is given by
\begin{equation}\label{jcr1}
j_x(x)=e\,\int_{p_{\rm l}}^{p_{\rm u}} \frac{c}{3}f_1\,4\,\pi\,p^2\,dp=e\,\int_{p_{\rm l}}^{p_{\rm u}}\frac{4\,\pi\,c}{3}F_1\,d\ln p,
\end{equation}
where $p_{\rm l}$ and $p_{\rm u}$ are, respectively, the lower and upper boundary momenta considered in the simulation that are set as $p_{\rm l}=0.1$ GeV and $p_{\rm u}=1$ PeV in this paper, and we have assumed that the cosmic-rays are composed of protons.
According to the detailed linear analysis by Bell (2004), the cosmic-rays whose gyroradius ($r_{\rm g}=p\,c/e\,B$) is smaller than an unstable scale does not contribute to the current inducing the non-resonant Bell instability.
This stems from the fact that such cosmic-rays induce a current perpendicular to the $x$-axis that weakens the Lorentz force driving the instability.
Hence, in order to obtain a realistic growth of the Bell instability, we use the following current density instead of eq.~(\ref{jcr1}):
\begin{equation}\label{jcr2}
j^{(\rm cr)}_x(x)=e\,\int_{p_{\rm B}}^{p_{\rm u}}\frac{4\,\pi\,c}{3}F_1\,d\ln p,
\end{equation}
where the lower bound of the integral $p_{\rm B}$ is determined by the condition $p_{\rm B}c/e\,B=l_{\rm B, min}$.
Here $l_{\rm B, min}=c\,B_x/4\pi\,j^{(\rm cr)}_x$ is the minimum scale of the Bell instability.
Simple algebra yields
\begin{equation}\label{ps}
p_{\rm B}=\frac{e\,B_{x}\,B}{4\pi\,j^{(\rm cr)}_x}.
\end{equation}
When $p_{\rm B}$ is not found in the range between $p_{\rm l}$ and $p_{\rm u}$, we set $j^{(\rm cr)}_x=0$.

We employ the following diffusion coefficient due to the pitch angle scattering (Skilling 1975)
\begin{equation}\label{kappa}
\kappa(p,{\bf B})=\frac{4}{3\,\pi}\frac{\max(B_x^2,\delta B^2)}{\delta B^2}\frac{v_{\rm CR}\,p_{\rm CR}\,c}{e\,\max(|B_x|,\delta B)},
\end{equation}
where $\delta B^2=B_y^2+B_z^2$.
In this expression, we made two assumptions (1) the magnetic field fluctuation $\delta B$ always contributes to the scattering of the particles regardless of the scale of the fluctuation, and (2) the Bohm limit diffusion estimated by using $\delta B$ is realized when the turbulent component dominates the ordered field $B_x$.
The former assumption is not problematic, at least in the present situation, because the typical scale of the fluctuation $\delta B$ induced by the Bell instability is on the order of the gyro radius of TeV particles, and a turbulent cascade would naturally provide scatterers for sub-TeV particles, which are most important for the gamma ray spectral modification (see \S \ref{RES} and \S 3).
The latter assumption is quite reasonable, although $\kappa$ does not reach the Bohm limit ($\delta B=B_x$) in the present simulation (see Figure \ref{f1}).

\subsection{Initial and Boundary Conditions}
In this paper, we examine the external irradiation of a uniform cold cloud by cosmic-rays at $t=0$, which approximately corresponds to the time when the SNR blast wave hits the cloud.
For this purpose, we initially set a uniform gas of density $\rho_{\rm c}=500\,m_{\rm p}$ g cm$^{-3}$ and temperature $T_{\rm c}=20$ K with no internal cosmic-rays ($f_0=f_1=0$), and set the cosmic-rays that irradiate the cloud at a boundary ($x=0$) as 
\begin{equation}
f_0(x=0,p)=f_{\rm ext}\,p^{-s}\,\exp(-p/p_{\rm max}).
\end{equation}
These incident cosmic-rays are assumed to be accelerated at the SNR shock wave before it hits the cloud, and hence we adopt the prediction of the standard DSA value of $s=4$.
The maximum momentum is fixed as $p_{\rm max}\,c=300$ TeV, which is suggested in the case of RX J1713.7$-$3946 by Gabici \& Aharonian (2014).
The normalization parameter $f_{\rm ext}$ depends on the acceleration efficiency of the DSA.
Given that a fraction $\zeta$ of the supernova energy $E_{\rm SN}=10^{51}$ erg is deposited as the cosmic-ray energy, it becomes
\begin{equation}\label{fext}
f_{\rm ext}=\frac{\zeta\,E_{\rm SN}}{4\,\pi\,c\,\log(p_{\rm max}/p_{\rm GeV})V_{\rm SNR}},
\end{equation}
where $V_{\rm SNR}$ is the volume of the SNR and we have used $s=4$ and the fact that the DSA deposits the energy for particles with energy above GeV.
We use $\zeta=0.1$ and set the radius of the SNR as $R=8$ pc, which gives $f_{\rm ext}=3.34\times10^{-22}$ erg s cm$^{-4}$.
For $f_1$, we use the free boundary condition at $x=0$.
Another spatial boundary is set at $x= L=2$ pc, where the free boundary condition is imposed for both $f_0$ and $f_1$.

For the initial magnetic field, we examine the two cases of $B_x=5$ and $10\,\mu$G, which are reasonable but slightly weaker than the typical magnetic field strength in molecular clouds (Crutcher et al. 2010).
As a seed of the Bell instability that induces the Alfv\'en waves, we initially put small $B_z$ and $B_y$ fluctuations as the white noise with a dispersion of $\langle \delta B^2\rangle^{1/2}=0.009\,B_x$.
Because the magnetic fields in molecular clouds are expected to be well ordered at least on a parsec scale (see, e.g., Inoue \& Inutsuka 2012 for simulation), and the gyroscale of the relativistic cosmic-rays, $\sim 7\times 10^{14}\,(E_{\rm CR}/1\,\mbox{TeV})\,(B/5\,\mu\mbox{G})^{-1}$ cm, is much smaller than the scale of the cloud, $\sim 10^{18}$ cm, the one-dimensional treatment of the cosmic-ray propagation along the $x$-axis is justified.

After the SNR blast wave hits, a shock wave is transmitted into the cloud.
In this paper, we neglect the effects of the shock propagation in the cloud for the following reasons:
Because of the high cloud density, the speed of the shock is stalled inside cloud, and hence it does not further accelerate the cosmic-rays.
The shock compression amplifies the magnetic field in addition to the Bell instability, which indicates from eq.~(\ref{kappa}) that the diffusion coefficient can be reduced by a factor of $1/r^2$ behind the shock, where $r$ is the compression ratio of the shock.
The diffusion length $l_{\rm d}$ is also reduced as $l_{\rm d}\propto \kappa^{1/2}\propto 1/r$.
Thus, because $l_{\rm d}\,\rho\propto r^0$, the amount of gas particles that interact with the cosmic-rays remains unchanged due to the shock propagation, implying that the effect of the shock propagation in the cloud would be limited for the gamma ray emission.

\subsection{Resolution}\label{RES}
As we shall show in the next section, the typical strength of the cosmic-ray current density in the cloud is $j^{(\rm{cr})}_x\sim 10^{-10}$ esu s$^{-1}$ cm$^{-2}$, which is mostly due to the streaming of particles above TeV energy
\footnote{
Under the $f_{\rm ext}$ value given by eq.~(\ref{fext}), the free streaming of cosmic-rays with energies above 1 TeV gives the current $j_x=7\times 10^{-10}$ esu s$^{-1}$ cm$^{-2}$.
The free streaming is possible only in very early stage and streaming velocity goes down as diffusion coefficient is enhanced by the growth of the Bell instability.
After 100 yr, the current becomes around $10^{-10}$ esu s$^{-1}$ cm$^{-2}$.
}.
Thus, the most unstable spatial and time scales of the Bell instability can be estimated as (Bell 2004)
\begin{eqnarray}
k_{\rm B}^{-1}&=&\frac{2c\,\rho_{\rm c} v_{\rm A, c}^2}{|j^{(\rm{ret})}_x| B_x}=\frac{c\,B_x}{2\pi\,|j^{(\rm{cr})}_x|} \nonumber\\
&\simeq& 2\times 10^{14} \mbox{ cm} \,\left( \frac{j^{({\rm cr})}}{10^{-10}\,\mbox{esu s$^{-1}$cm$^{-2}$}} \right)^{-1}\left( \frac{B_x}{5\,\mu\mbox{G}} \right), \label{kB}\\
\omega_{\rm B}^{-1}&=&\frac{1}{k_{\rm B}\,v_{\rm A, c}}=\frac{c\,\rho_{\rm c}^{1/2}}{\pi^{1/2}\,|j^{(\rm{cr})}_x|} \nonumber\\
&\simeq& 150 \mbox{ yr} \,\left( \frac{j^{({\rm cr})}}{10^{-10}\,\mbox{esu s$^{-1}$cm$^{-2}$}} \right)^{-1}\left( \frac{\rho_{\rm c}}{500\,m_{\rm p}\mbox{g cm}^{-3}} \right)^{1/2}. \label{tB}
\end{eqnarray}
In order to resolve this scale with more than 10 numerical cells, we need a numerical cell number $N_{\rm cell}\gtrsim L/(0.1\,k_{\rm B}^{-1})\sim 10^5$ indicating that a very high resolution is required.
To satisfy this requirement, we use $N_{\rm cell}=2^{19}=262144$ ($\Delta x=L/N_{\rm cell}=2.4\times 10^{13}$ cm).
For the momentum space, we consider the range $p_{\rm l}\,c=10^8\mbox{ eV}\le p\,c \le p_{\rm u}\,c=10^{15}\mbox{ eV}$, which is divided into uniform 128 numerical cells in the logarithmic scale, i.e., $\Delta \ln p=\ln (10^{7})/128$.

\subsection{Numerical Schemes}
The MHD equations (\ref{eq1})-(\ref{eq8}) except the Bell term (RHS of [\ref{eq3}] and [\ref{eq4}]) are integrated using the second-order Godunov-type scheme with an approximate Riemann solver developed by Sano et al.~(1999).
The Bell terms are solved by using the piecewise exact solution (PES) method developed by Inoue \& Inutsuka (2008). 
The telegrapher-type diffusion convection equations (\ref{DC1}) and (\ref{DC2}) are integrated using the fourth-order MUSCL scheme (Yamamoto \& Daiguji 1993), except the second term of the RHS of eq.~(\ref{DC2}), which is integrated by the PES method.
Because the PES method is combined with a second-order operator splitting technique, the system as a whole is consequently solved with the second-order accuracy.

The timestep of integration is determined by $\Delta t=\min[ \Delta t_{\rm fs}, \Delta t_{\rm ah}]$, where $\Delta t_{\rm fs}=c_{\rm cfl}\,\Delta x/(c/\sqrt{3})$ is the Courant-Friedrichs-Lewy (CFL) condition for the free streaming cosmic-rays, and $\Delta t_{\rm ah}=3\,(\partial v_x/\partial x)^{-1}\,\Delta \log p\,c_{\rm cfl}$ is the CFL condition for the adiabatic cosmic-ray heating term.
The CFL number $c_{\rm cfl}=0.5$ is used in the present runs.
Note that we do not need other timestep limiters, because the characteristic velocity from the MHD part hardly exceeds $c/\sqrt{3}$, and the PES does not impose a time-step limitation.

\subsection{Advantage of Telegrapher Type Equations} \label{adv}
One can perform a similar simulation by solving the usual diffusion convection equation, in which the r.h.s.~of equation (\ref{DC1}) should be replaced by the diffusion term $\partial_x\,\{\kappa\,(\partial_x\,F_0)\}$.
In such an equation, we have to impose the stability condition $\Delta t\le\Delta t_{\rm d}=\Delta x^2/(2\, \kappa)$ instead of $\Delta t_{\rm fs}$.
In the case of the present numerical setting, $\Delta t_{\rm d}\ll 10^{-2}$ s for PeV particles, while $\Delta t_{\rm fs}\simeq 700$ s, indicating that the use of the telegrapher-type equations is computationally quite advantageous.
Note that we could use a larger timestep for the usual diffusion convection equation than $t_{\rm fs}$ if we employ an implicit scheme.
However, in general, the first order implicit scheme is not reliable for a dynamical problem in accuracy and higher order implicit schemes are usually problematic in terms of their numerical stability.

We integrate the basic equations for $400$ years that requires $\simeq 2\times 10^{7}$ time step.
To perform this calculation, we employ parallel supercomputer Cray XC30 and XC50 systems, whose cost is approximately 160,000 CPU hours for one run.

\subsection{Ion-neutral Friction Damping}\label{AD}
In molecular clouds, ion-neutral collisional friction is an important agent that damps out the Alfv\'en waves.
The timescale of the wave damping for the Alfv\'en waves at the most unstable scale of the Bell instability can be estimated as (Braginskii 1965; Kulsrud \& Pearce 1969):
\begin{eqnarray}
t_{\rm in}&\simeq&\frac{2\,\nu_{\rm in}\,x_{\rm i}}{k_{\rm B}^2\,v_{\rm A,c}^2}=\frac{2\,\nu_{\rm in}\,x_{\rm i}\,c^2\,\rho}{\pi\,|j^{(\rm{cr})}_{x}|^2} \nonumber\\
&\simeq& 1500\,\mbox{yr}\,\left( \frac{x_{\rm i}}{10^{-3}} \right)\left( \frac{\rho}{500\,m_{\rm p}\,\mbox{g cm}^{-3}} \right)^{2}\left( \frac{j^{({\rm cr})}}{10^{-10}\,\mbox{esu s$^{-1}$cm$^{-2}$}} \right)^{-2},
\end{eqnarray}
where $\nu_{\rm in}=2\times 10^{-9}\,n_{\rm c}$ s$^{-1}$ is the momentum exchange frequency for an ion in a field of neutrals (Osterbrock 1961), and $x_{\rm i}$ is the ionization fraction.
The estimation indicates that the timescale of the wave damping can be longer than that of the Bell instability as long as $x_{\rm i}\gtrsim10^{-4}$.
Although the ionization fraction of the molecular cloud in the vicinity of a young SNR is unknown, the ionization fraction in an optically thin cold cloud is calculated to be $x_{\rm i}\sim 10^{-3.5}$ for the gas of $n\sim 10^{2-4}$ cm$^{-3}$ (e.g., Koyama \& Inutsuka 2000)
\footnote{For the spectral modification, we need growth of the Bell instability only at the surface region of the cloud. This is the reason why we apply optically thin calculation, rather than that of optically thick cloud.}.

Reville et al.~(2007) showed that the growth timescale of the Bell instability in a partially ionized medium is approximately given by eq.~(\ref{tB}) for the case $\nu_{\rm ni}=x_{\rm i}\,\nu_{\rm in}> \omega_{\rm B}$ (one can verify the growth timescale by solving eq.~(13) of Reville et al.~2007, bearing in mind that $v_{\rm A}$ in Reville et al. 2007 is defined as $B/\sqrt{4 \pi \rho_{\rm i}}$).
This condition ($x_{\rm i}\,\nu_{\rm in}> \omega_{\rm B}$) can be rewritten as $x_{\rm i}> 10^{-4}$ in the present case, which seems to be reasonable ionization degree for cloud surface regions.

In the vicinity of a young SNR, the ionization degree can be much larger than the typical ISM value, because of the enhanced x-ray ionization due to nonthermal synchrotron emissions.
In addition, after the cosmic-ray irradiation, we can expect stronger cosmic-ray ionization.
Furthermore, after shock sweeping, because the temperature of cloud rises drastically, the damping timescale would be prolonged more than $10^3$ yr.
Therefore, in this paper, we omit the effect of the ion-neutral friction wave damping.
Note, however, that if we consider dynamics in the inner region of the cloud where $x_{\rm i}<\omega_{\rm B}/\nu_{\rm in}\simeq10^{-4}$ or longer timescale dynamics, we need to consider the effect of the ion-neutral friction seriously.
According to Reville et al.~(2007), even if ionization degree is smaller than $\omega_{\rm B}/\nu_{\rm in}$, the friction does not stabilize the Bell instability, but the growth rate will depend explicitly on the $\nu_{\rm in}$ and $x_{\rm i}$ (eq.~[16] of Reville et al.~2007).
The influences of the ion-neutral friction on the growth rate of the instability and the damping rate of induced waves can be treated if we consider the ion-neutral two-fluid system (see, e.g., Inoue et al. 2007 for full set of two-fluid equations) instead of using the strong coupling limit applied in this present paper.

\section{Results and Implications for Gamma Ray Emissions}
\subsection{Results of the $B_x=5\,\mu$G Case}
\begin{figure*}[h]
\includegraphics[scale=1.2]{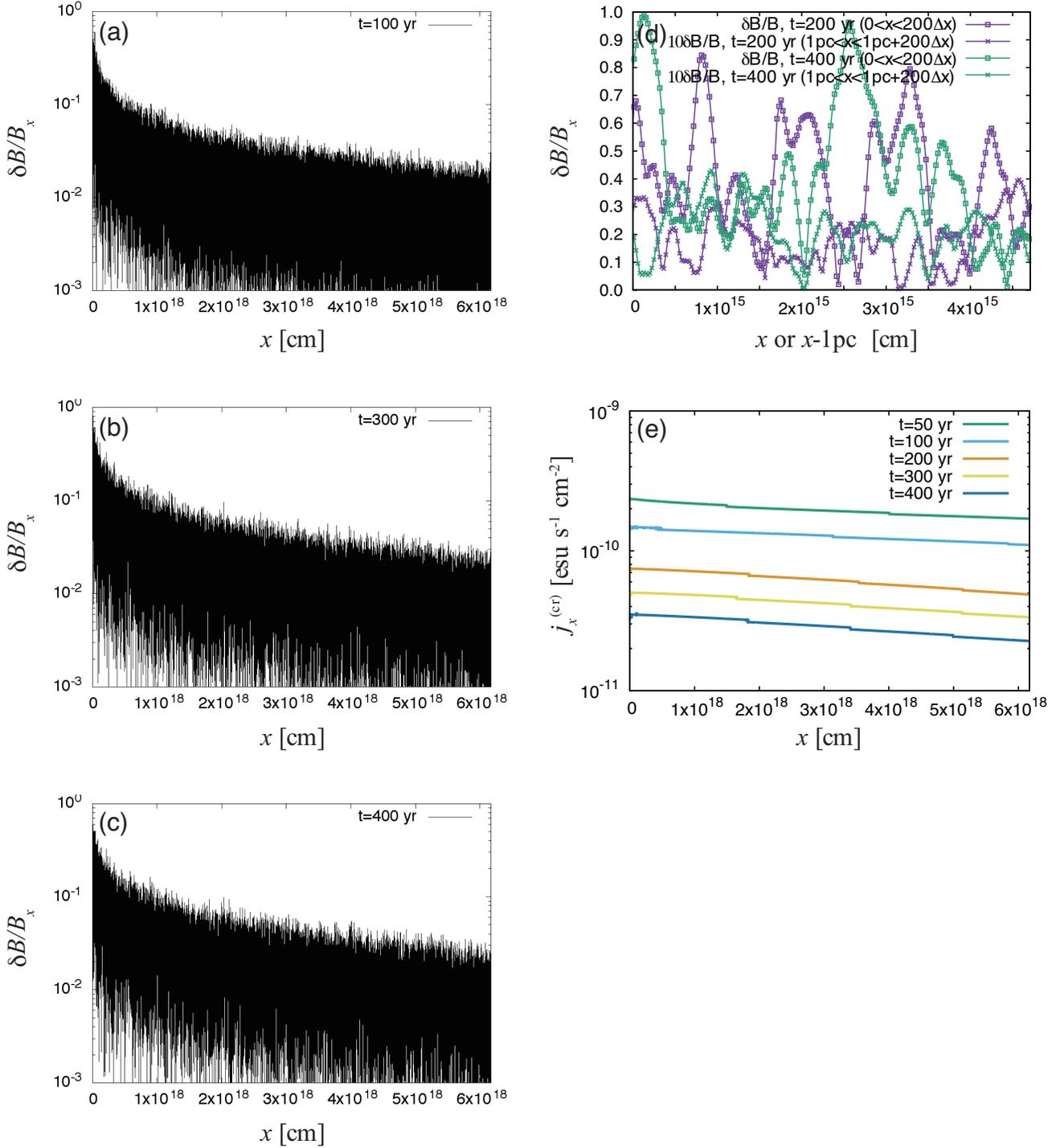}
\caption{\label{f1}
Structure of the magnetic field and the cosmic-ray current density.
Panels (a)-(c) represent the degree of magnetic field disturbances $\delta B/B_x$ at $t=100,\,300,$ and $400$ yr, respectively.
In Panel (d), detailed structures of $\delta B/B_x$ are plotted to show that turbulent structures are well resolved, where squares and crosses are the structures in the region $0\le x\le200\,\Delta x$ and $1\mbox{ pc}\le x\le 1\mbox{ pc}+200\,\Delta x$, respectively, and different color represents different time of snapshot.
In Panel (e), the structure of the cosmic-ray current density $j^{(\rm{cr})}_x$ is plotted.
}\end{figure*}

We first show the results of the run with $B_x=5\,\mu$G.
The return current due to the cosmic-ray streaming in the cloud induces the Bell instability that creates (circularly polarized) Alfv\'en waves.
In Figure \ref{f1} we show the structure of the magnetic field and the cosmic-ray current density.
Panels (a)-(c) represent the degree of magnetic field disturbances $\delta B/B_x$ at $t=100,\,300,$ and $400$ years, respectively.
In Panel (d), detailed structures of $\delta B/B_x$ are plotted to demonstrate that turbulent structures are well resolved, where squares and crosses are, respectively, the structures in the region $0\le x\le200\,\Delta x$ and in the region $1\mbox{ pc}\le x\le 1\mbox{ pc}+200\,\Delta x$, and different color represents different time of snapshot.
In Panel (e), the structure of the cosmic-ray current density $j^{(\rm{cr})}_x$ is plotted.
Because the generated Alfv\'en waves reduce the diffusion coefficient, particles with energies $E\lesssim 1$ TeV are dammed in the shallow region of the cloud.
The higher the particle energy, the deeper the penetration depth, as a result of which the shallower region has a stronger current and thus amplitude of the Alfv\'en waves.

Unlike the usual Bell instability simulations in a shock upstream (see, e.g., Riquelme \& Spitkovsky 2009), the amplitude of the Alfv\'en waves does not enter the nonlinear regime ($\delta B>B_x$).
This stems from the fact that, in the present simulation, the cosmic-ray current decreases with time because the induced Alfv\'en waves work to prevent further cosmic-ray incursion.

\begin{figure}[t]
\includegraphics[scale=0.65]{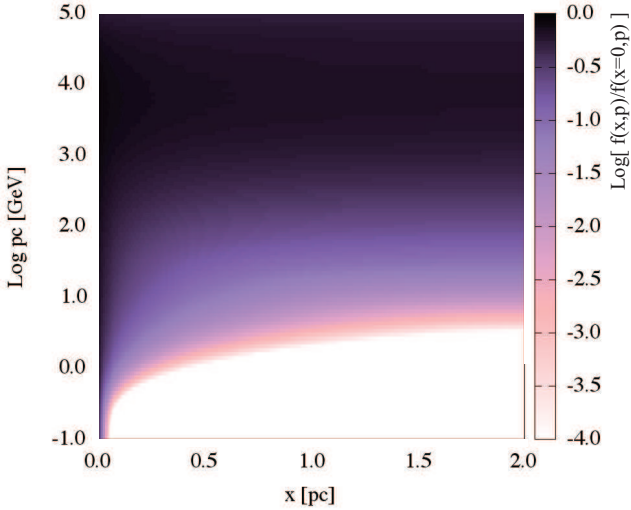}
\caption{\label{f2}
Spectral cosmic-ray density normalized by the external (or boundary) value: $f(x,p)/f(x=0,p)$ at $t=400$ yr.
}\end{figure}

\begin{figure}[h]
\includegraphics[scale=0.65]{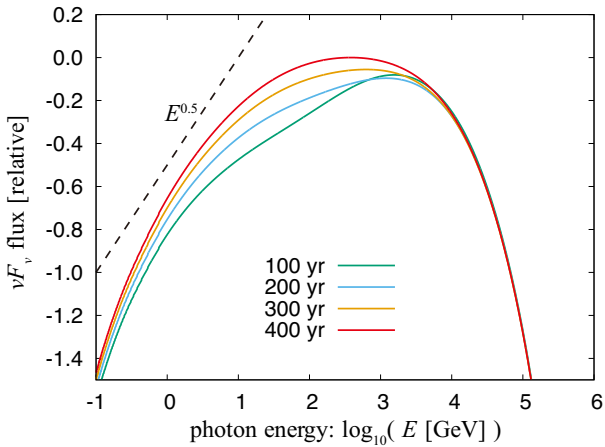}
\caption{\label{f3}
Synthetic gamma ray $\nu\,F_{\nu}$ spectra using the cosmic-ray abundance inside the cloud $N(p)=\int f\,dx$ at $t=100$ yr (green), 200 yr (light blue), 300 yr (orange) and 400 yr (red).
The dashed line is proportional to $E^{0.5}$.
}\end{figure}

To see the energy-dependent diffusion, we plot the spectral cosmic-ray density normalized by the external value $f(x,p)/f_{0}(x=0,p)$ at $t=400$ yr in Figure \ref{f2}.
We see that only the particles with energy $E\gtrsim 1$ TeV can fully penetrate the parsec-scale cloud in the timescale of young SNRs $\lesssim 1000$ yr.
The slower diffusion for the sub-TeV energy particles affects the hadronic gamma ray spectrum.
In Figure \ref{f3}, we plot synthetic gamma ray $\nu\,F_{\nu}$ spectra based on the cosmic-ray abundance inside the cloud $N(p)=\int f_0\,dx$, by employing the formula given by Naito \& Takahara (1994) and Kamae et al.~(2006).
Green, blue, yellow, and red lines, respectively, show the spectra at $t=100,\,200,\,300$ and $400$ yr.
The spectral shape is similar to the gamma ray emission from the SNR RX J$1713.7-3946$, because it is approximately proportional to $E^{0.5}$ around $\sim10$ GeV (Abdo et al. 2011).
It has already been shown that the total hadronic gamma ray flux can be matched to the observations if the total mass of the cloud is $\sim 500\,M_{\odot}$ and $\zeta=0.1$ by Gabici \& Aharonian (2014).
In Inoue et al.~(2012), it is analytically shown that we can obtain the $\nu\,F_{\nu}$ spectral index $\simeq 0.5$ when we consider cosmic-ray propagation under the diffusion coefficient given by eq.~(\ref{kappa}). 

\begin{figure}[t]
\includegraphics[scale=0.65]{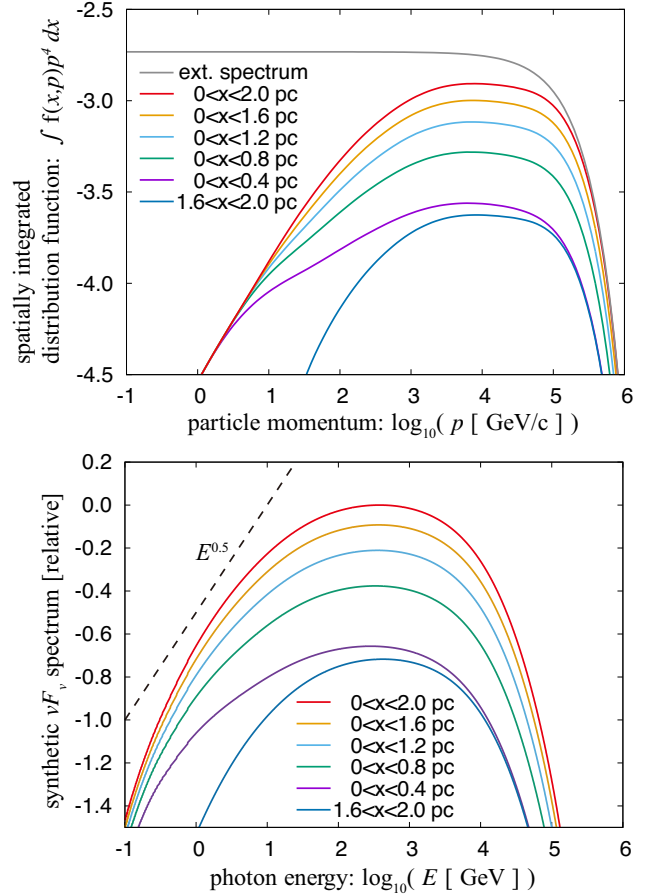}
\caption{\label{f4}
Cosmic-ray spectra and synthetic gamma ray $\nu\,F_{\nu}$ spectra constructed from various spatial regions.
The top panel shows the cosmic-ray spectra obtained by integrating the distribution function in the spatial ranges specified in the legend: $\int f(x,p)\,p^4\,dx$.
The bottom panel exhibits the synthetic $\nu\,F_{\nu}$ gamma ray spectra based on the cosmic-ray spectra shown in the top panel.
}\end{figure}

Our result suggests that the gamma ray spectrum varies its shape depending on the depth of the cloud to which the gamma rays are emitted.
For instance, in Figure \ref{f4}, we plot cosmic-ray spectra and synthetic gamma ray $\nu\,F_{\nu}$ spectra constructed from various spatial regions.
The top panel shows the cosmic-ray spectra that are obtained by integrating the distribution function in the spatial ranges specified in the legend $\int f(x,p)\,p^4\,dx$.
The bottom panel exhibits the synthetic $\nu\,F_{\nu}$ gamma ray spectra based on the cosmic-ray spectra shown in the top panel.
Because particles with higher energy have larger diffusion lengths, the $\nu\,F_{\nu}$ spectrum of the gamma rays from the shallow region (magenta) is flatter than those of the deep regions (e.g., blue).

\subsection{Results of the $B_x=10\,\mu$G Case}
\begin{figure}[t]
\vspace{-1.cm}
\includegraphics[scale=0.86]{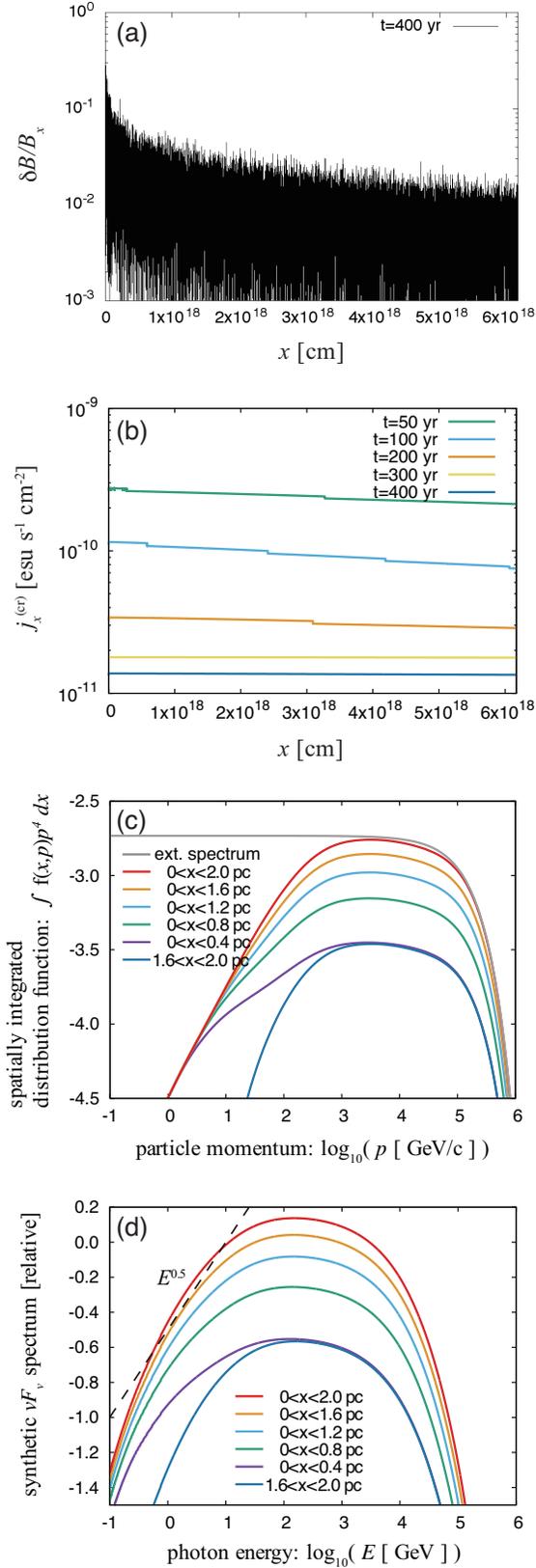}
\caption{\label{f5}
Panel (a): Structure of the magnetic field at $t=400$ yr.
Panel (b): Structure of the cosmic-ray current density in various epochs.
Panel (c): Cosmic-ray spectra obtained by integrating the distribution function in the spatial ranges specified in the legend: $\int f(x,p)\,p^4\,dx$.
Panel (d): Synthetic $\nu\,F_{\nu}$ gamma ray spectra based on the cosmic-ray spectra shown in the top panel.
}\end{figure}

For the case $B_x=10\,\mu$G, we obtained results that are very similar to the $B_x=5\,\mu$G case.
In Figure \ref{f5}, we show the structures of the magnetic field (panel [a]) and the cosmic-ray current density (panel [b]), and the spectra of the cosmic-rays (panel [c]) and the synthetic gamma rays (panel [d]).
The main difference compared to the $B_x=5\,\mu$G case is that the cosmic-rays penetrate further into the cloud.
The reason for this somewhat nontrivial result is explained as follows:
Because of the larger $B_x$, the minimum momentum of the cosmic-rays that can contribute to the non-resonant instability $p_{\rm B}$ becomes smaller (see, eq.~[\ref{ps}]).
This leads to a smaller current density, and thus results in a less turbulent and smaller $\kappa$ medium for the cosmic-rays.

\subsection{Convergence Check}
\begin{figure}[t]
\vspace{0.cm}
\includegraphics[scale=0.65]{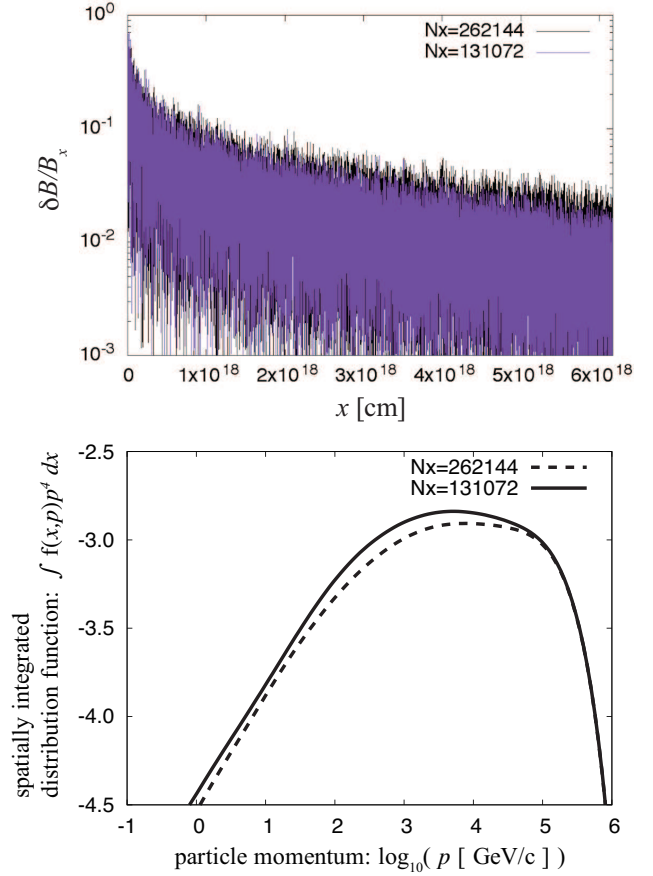}
\caption{\label{f6}
Top: Structure of the magnetic field fluctuations at $t=400$ yr.
Black line corresponds to the result of the fiducial run and purple line corresponds to the lower resolution one.
Bottom: Spatially integrated cosmic-ray spectrum: $\int f(x,p)\,p^4\,dx$.
Dashed line corresponds to the result of the fiducial run and solid line corresponds to the lower resolution one.
}\end{figure}

To check numerical convergence, we execute an additional simulation for $B_x=5\,\mu$G case with the half spatial resolution ($N_{\rm cell}=131072$).
In Figure \ref{f6}, the resulting spatial structure of the magnetic field fluctuations ({\it top}) and the cosmic-ray spectrum ({\it bottom}) at $t=400$ yr are shown with the fiducial resolution results ($N_{\rm cell}=262144$).
We see very reasonable results that the lower resolution run exhibit a bit more dissipative magnetic field structure (slightly lower amplitude of $\delta B/B_x$ than the fiducial result), and the cosmic-rays penetrate a bit more into the cloud than the fiducial one.
The overall differences are not substantial, so we can say that the resolution of our fiducial run would be enough to make a conclusion.

\section{Summary and Discussion}
We have studied the spectral modification of hadronic gamma rays due to the turbulent magnetic field induced by the Bell instability in a molecular cloud interacting with cosmic-rays accelerated at a young SNR shock.
In order to examine a cosmic-ray incursion into a cloud, we solved a hybrid system of the Bell MHD equations and the telegrapher type diffusion convection equations (eqs.~[\ref{eq1}]-[\ref{DC2}]).
We have shown that, at least in the present parameter set, the Bell instability successfully induces a turbulent magnetic field that prevents the incursion of the cosmic-rays with energy $E\lesssim 1$ TeV into the cloud.
The synthetic hadronic gamma ray spectrum resembles the observed gamma ray spectrum.
Our result predicts that the $\nu\,F_{\nu}$ spectrum of the gamma rays from the shallow region of the cloud are flatter than those of the deep region.
This prediction can be proved if we have a gamma ray telescope that has a sub-parsec spatial resolution in the GeV to TeV energy range.
In the case of SNR RX J$1713.7-3946$ ($D\sim 1$ kpc; Fukui et al. 2003), the required resolution is approximately 100 arcsec., which can be achieved by the Cherenkov Telescope Array in the TeV range but is difficult to achieve with any instrument in the GeV range.

Finally, we wrap up this paper by pointing out the potential flaws of our calculation.
We have assumed that the turbulent magnetic field always contributes to the scattering of the cosmic-rays regardless of their energy.
The most unstable scale of the Bell instability is estimated in eq.~(\ref{kB}) that is slightly larger than the gyro radii of sub-TeV particles.
Thus if we consider a cascade of turbulence, we can naturally expect efficient scattering for sub-TeV particles, which are the most important for spectral modulation.
However, in the case of sub-Alfv\'enic MHD turbulence, it is said that the turbulent cascade does not efficiently create smaller-scale Alfv\'en waves (Goldreich \& Sridhar 1995, Yan \& Lazarian 2002).
Nevertheless, it is possible to anticipate efficient scattering, because the Bell instability selectively induces circularly polarized Alfv\'en waves, which are known to be unstable and create daughter waves (Goldstein 1978).
To confirm this expectation we need a further high-resolution simulation that can follow the cascade due to the instability, which will be attempted in our future works.

The second potential flaw is from our neglect of ion-neutral frictional damping.
In \S \ref{AD}, we omitted it, because its timescale can be longer than the growth timescale of the Bell instability.
To confirm whether we can really neglect the effect of the wave damping, we need to calculate the detailed degree of ionization in the cloud by taking into account the ionization by the high-energy cosmic-ray incursion and shock heating.

\acknowledgments
T.I. is grateful to S. Gabici for a very insightful discussion, which provided a basic idea of this work.
T.I. also thanks F. Takahara, T. Terasawa, R. Yamazaki and Y. Ohira for fruitful comments.
The numerical computations were carried out on XC30 and XC50 system at the Center for Computational Astrophysics (CfCA) of National Astronomical Observatory of Japan.
This work is supported by Grant-in-aids from the Ministry of Education, Culture, Sports, Science, and Technology (MEXT) of Japan (15K05039, 18H05436).

\end{document}